\documentclass[aps,prl,twocolumn,floats,letterpaper,superscriptaddress,tighten]{revtex4}
\usepackage{amssymb}
\usepackage{amsbsy}
\usepackage{amsmath}
\usepackage{graphicx}
\usepackage{graphics}
\usepackage{setspace}
\usepackage{subfigure}
\usepackage{array}
\usepackage{color}
\usepackage{fontenc}
\usepackage{textcomp}
\usepackage{rotating}
\usepackage{bm}
\usepackage{hyperref}
\hypersetup{pdfpagemode=UseNone}

\newcommand{\bsub}{\begin{subequations}}
\newcommand{\esub}{\end{subequations}}
\newcommand{\vex}[1]{\bm{\mathrm{#1}}}
\DeclareMathOperator{\Tr}{Tr}
 
\newcommand{\rcl}{\rho_{\mathsf{cl}}}
\newcommand{\QSP}{Q_{\mathsf{SP}}}
\newcommand{\Gcl}{G_{\mathsf{cl}}}

\newcommand{\bPsi}{\bar{\Psi}}

\newcommand{\kb}{\vex{k}}
\newcommand{\qb}{\vex{q}}
\newcommand{\bl}{\vex{l}}
\newcommand{\rb}{\vex{r}}

\newcommand{\braket}[1]{\left\langle {#1} \right\rangle}
\newcommand{\Avert}[1]{\lvert {#1} \rvert}
\newcommand{\ket}[1]{\left| {#1} \right\rangle}
\newcommand{\bra}[1]{\left\langle {#1} \right|}

\newcommand{\intl}[1]{\int\limits_{#1}}

\newcommand{\lb}{\lambda}
\newcommand{\dd}{\mathcal{D}}
\newcommand{\ww}{\omega}
\newcommand{\bd}{\Delta}

\newcommand{\msf}[1]{\mathsf{#1}}

\begin{document}

\title{Dephasing catastrophe in $4 - \epsilon$ dimensions:\\
	A possible instability of the ergodic (many-body-delocalized) phase}
\author{Yunxiang Liao}
\affiliation{Department of Physics and Astronomy, Rice University, Houston, Texas 77005, USA}
\author{Matthew S.\ Foster}
\affiliation{Department of Physics and Astronomy, Rice University, Houston, Texas 77005, USA}
\affiliation{Rice Center for Quantum Materials, Rice University, Houston, Texas 77005, USA}
\date{\today\\}


\begin{abstract}
In two dimensions (2D), dephasing by a bath cuts off Anderson localization that would otherwise occur 
at any energy density for fermions with disorder. 
For an isolated system with short-range interactions, the system can be its own bath, 
exhibiting diffusive (non-Markovian) thermal density fluctuations. 
We recast the dephasing of weak localization due to a diffusive bath
as a self-interacting polymer loop.
We investigate the critical behavior of the loop in $d=4-\epsilon$ dimensions, and find a nontrivial 
fixed point corresponding to a temperature $T^* \sim \epsilon >0$ where the dephasing time \emph{diverges}. 
Assuming that this fixed point survives to $\epsilon=2$, we associate it 
to a possible instability of the ergodic phase.
Our approach may open a new line of attack against the problem of the ergodic to many-body-localized phase 
transition in $d > 1$ spatial dimensions. 
\end{abstract}

\maketitle


\textit{Introduction}.---
The interplay between quantum interference  
and inelastic quasiparticle scattering in a disordered medium  
takes center stage 
in the problem of many-body localization (MBL) \cite{BAA,BAA2,Gornyi05,MBL-Rev}. 
Although the MBL phase~\cite{Imbrie2016} and the ergodic metal-to-MBL insulator transition~\cite{MBL1DT1,MBL1DT2,MBL1DT3} 
have been explored extensively in $d=1$ spatial dimension, their nature or even existence in 
higher dimensions remain open questions \cite{RH2017}.

Instead of the many-body localized phase, in this work 
we reconsider the \emph{standard theory of the ergodic phase} \cite{AAK,AA,AAG} in $d > 1$ spatial dimensions \cite{SuppMaterial}.
We identify a ``hole'' in this theory, when applied to a system that could transition 
to the MBL phase at low temperature. The hole concerns dephasing, which stabilizes the ergodic phase at finite temperature in 2D for 
an isolated system of fermions with weak disorder and short-range interactions.
We show that calculating the dephasing rate 
due to short-ranged interactions 
for the first quantum correction to transport (weak localization) 
is tantamount to computing a certain correlation function in a strongly coupled, 
auxiliary quantum field theory (QFT). 
While there exists a standard result for this case (e.g., \cite{Zala-Deph}), it is in fact 
a mean-field approximation (the self-consistent Born approximation SCBA), and mean field theory is expected to be 
unreliable for any 
field theory below its upper critical dimension \cite{CardyBook,GoldenfeldBook}. 
Within a controlled $\epsilon$-expansion, we identify a nontrivial fixed point corresponding to a nonzero critical 
temperature at which the dephasing of weak localization appears to fail.
This hints at the possibility of describing the ergodic-to-MBL phase transition by approaching
from the ergodic side.

\begin{figure}[b!]
	\includegraphics[angle=0,width=.3\textwidth]{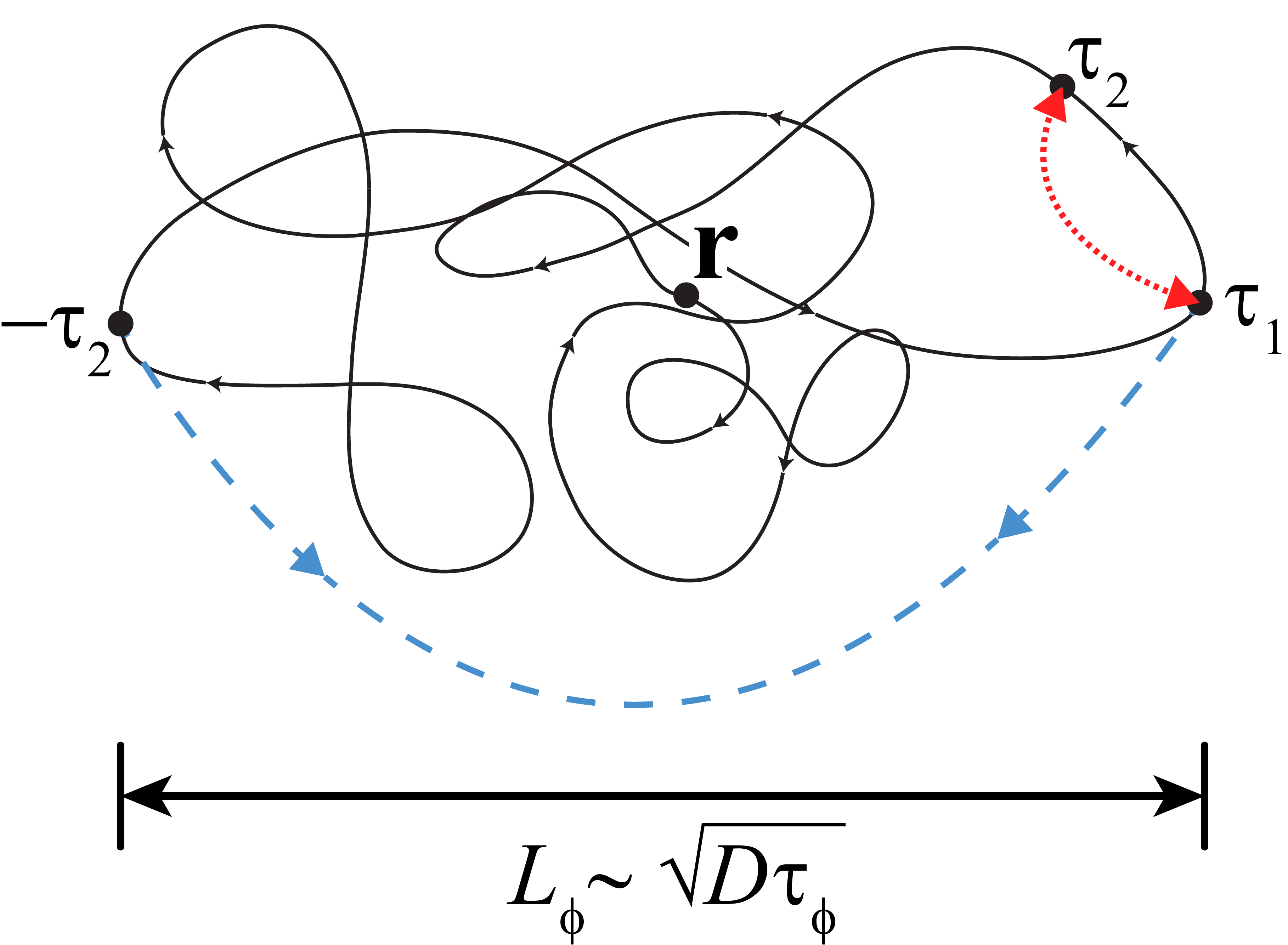}
	\caption{
		Cooperon as a self-interacting polymer loop: 
		The Cooperon $C_{\eta,-\eta}^{t}(\rb,\rb)$ entering the expression for the weak localization correction 
		[Eq.~(\ref{eq:AAK_S})] is given by the return probability of a diffusing particle interacting with a fluctuating density field bath
		$\rcl(\vex{r},t)$ \cite{AAK,AA,AAG,AleinerBlanter,Keldysh,ChakravartySchmid}. 
		The diffusion process starts and ends at the same point $\rb$ from time $-\eta$ to $\eta$. 
		For short-range interactions, the density fluctuations are diffusive and, similar to the virtual diffusion process, 
		controlled by the diffusion constant $D$ [see Eq.~(\ref{eq:Gcl})].
		The fluctuation-averaged Cooperon (return probability) in Eq.~(\ref{eq:PathSol}) can be interpreted as the path integral of a 
		self-interacting polymer loop. The action is given by Eqs.~(\ref{eq:S0}) and~(\ref{eq:SI}),
		where $S_0$ is the action for the unperturbed random walk, and $S_I$
		is the self-interaction term.
		The first term in $S_I$ describes the repulsive (causal) interaction between $\rb(\tau_1)$ and $\rb(\tau_2)$ of range $\sqrt{D_c \lvert \tau_1-\tau_2 \rvert}$, 
		and is indicated by the red dotted line in the figure;
		the second term, denoted by the blue dashed line, represents the attractive (anticausal) 
		interaction between $\rb(\tau_1)$ and $\rb(-\tau_2)$ of the same range.	
		The intrinsic length scale of the polymer loop is determined by the dephasing length 
		$L_{\phi} = \sqrt{D\tau_{\phi}}$, 
		which should diverge as the system approaches the MBL transition point from the ergodic phase \cite{BAA}.
	}
	\label{fig:F0}
\end{figure}

We emphasize that we consider \emph{weak} disorder and ``order one'' strength interactions, similar to the standard literature \cite{AAK,AA,AAG} but 
different from Basko, Aleiner, and Altshuler \cite{BAA} (who focused on strong disorder and weak interactions). 
The dephasing problem identified and treated in this work does not arise in 
the theory of diffusive electrons in solid state materials, owing to the long-ranged nature of the Coulomb interaction. 
In the latter case, the Markovian character of the (approximate) dephasing kernel admits an exact solution \cite{AAK,AleinerBlanter}, 
which happens to be the same as the SCBA \cite{ChakravartySchmid}. 
 
The problem considered here is different from that of a system with well-localized single particle states interacting with an external bath. 
In that case there are arguments (see e.g.\ \cite{BAA,SG-1,SG-2}) that any bath with a continuous spectrum leads to thermalization. 
We treat the weak localization correction and its dephasing on the same footing, within the hydrodynamic framework \cite{AAK,AA,AAG,Keldysh} that describes 
how quantum interference corrections to transport are dephased \emph{within the same system}.

For disordered, interacting fermions in the ergodic phase at finite energy density (temperature), 
the weak localization correction to the conductivity, which results from the quantum interference of 
wave functions scattered by impurities, is cut off in the infrared by dephasing~\cite{AAK,AA,AAG}. 
The dephasing problem is equivalent to a virtual random walk 
(Cooperon return probability)
that interacts with a stochastic bath.
At low temperatures in a metal, the system serves as its own bath due to inelastic electron-electron collisions,
mediated by screened Coulomb interactions. In this case the bath is Ohmic, i.e.\ the fluctuations in time 
are Markovian for the relevant frequency range $\omega \lesssim k_B T$. 
The Markovian case is exactly solvable and gives a finite dephasing rate $1/\tau_\phi > 0 $ for any $T > 0$ \cite{AAK}. 
In contrast to the long-range Coulomb interaction~\cite{AAK, AleinerBlanter}, 
the self-generated bath of a fermion system with short-range interactions is \emph{non-Markovian} (diffusive) \cite{Keldysh},
and does not admit an exact solution. Since in this case the thermal fluctuations of the diffusive bath are slow, these 
could prove ineffective at amputating quantum interference corrections for sufficiently small $T$ and/or small diffusion constant 
$D$. 

If the dephasing rate vanishes, the weak localization correction diverges logarithmically in the infrared for an infinite system in 2D. 
The system is therefore completely localized and unable to act as a heat bath for itself.
This suggests the possibility to access the MBL-ergodic transition in 2D by approaching it from the metallic side. 
The transition could be explored as a ``dephasing catastrophe'' in a model with short-range interactions; the latter are 
believed to be a requirement for MBL \cite{Yao14}. 
In particular, if there exists a 2D system with a many-body mobility edge at a finite energy density corresponding to 
temperature $T_{\msf{MBL}}$, the dephasing of quantum interference corrections would fail as $T$ approaches $T_{\msf{MBL}}$ from above.

In this Letter, we consider the lowest order weak localization correction due to the virtual
return probability of the Cooperon. 
We recast the dephasing of the Cooperon due to a diffusive bath as a geometric 
statistical-mechanics problem of a self-interacting polymer loop, 
analogous to the self-avoiding random walk (SAW)---see Fig.~\ref{fig:F0}.
The self-interactions are themselves diffusive, as mediated by the bath. 
As with the SAW, we construct a replica field theory \cite{CardyBook}
whose upper critical dimension $d_c$ is $4$, and investigate the critical behavior 
using the renormalization group (RG) approach. 

In addition to the Gaussian fixed point (corresponding to decoupled Cooperon and bath),
we identify a nontrivial fixed point controlled by an $\epsilon$-expansion
with $\epsilon \equiv d_c - d > 0$. 
The fixed point has only one relevant direction, and  
its location corresponds to a finite temperature $T^* \sim \epsilon > 0$ 
at which the dephasing rate (``mass of the Cooperon'') vanishes.
We also compute the correlation length critical exponent $\nu$ to one-loop level, 
and find in $d$ spatial dimensions $\nu=2/d$, saturating the Harris-Chayes bound~\cite{Harris, Chayes}. 

As with all studies of this type \cite{CardyBook,GoldenfeldBook,ZinnJustinBook}, we cannot be sure that the fixed point 
we find survives to finite $\epsilon = 2$, which would be the relevant dimension for demonstrating an instability of the ergodic phase.
However, like all $\epsilon$-expansions we \emph{can} be sure that our result is not invalidated by the next order in perturbation theory, 
at least for sufficiently small $\epsilon \ll 1$. Moreover, we show that the dephasing problem represents a type of geometric criticality, 
and as such is governed by a nonunitary QFT. Critical points in nonunitary QFTs can arise even at the ``lower critical dimension'' 
($\epsilon = 2$), as is well-known for the self-avoiding walk \cite{ZinnJustinBook,CFT-Book} and weak antilocalization in the symplectic class \cite{AndRev}. 
In 3D ($\epsilon = 1$), our result predicts a kink in the temperature-dependence of the conductivity for an isolated, weakly disordered fermion system
with short-ranged interactions, since the infrared part of the weak localization correction is analytic in the inverse dephasing length. 

While our result suggests a possible instability of the ergodic phase in 2D, we cannot immediately identify this
with the ergodic-MBL transition.  It is possible that the MBL phase itself does not exist for $d > 1$ \cite{RH2017}. 
Moreover, at the ergodic-MBL transition, one must account for higher order 
quantum interference corrections; we return to this issue in the conclusion. 
We note however that our results should be testable via a classical lattice polymer simulation 
in two and three dimensions.


\textit{The problem}.---
The  weak localization (WL) correction to the conductivity, 
which is caused by the quantum interference between pairs of time-reversed paths, 
can be written as
\begin{align}
\label{eq:AAK_S}
	\delta \sigma_{\msf{WL}} 
	= 
	&
	-
	\frac{4 e^2}{\hbar \pi} D
	\intl{\eta}  
	\braket{C_{\eta,-\eta}^{t}(\rb,\rb)}_{\rho},	
\end{align}
where $C_{\eta,-\eta}^{t}(\rb,\rb)$ is the Cooperon satisfying the differential equation~\cite{AAK, Keldysh}:
\begin{align}\label{eq:AAK_C}
\begin{aligned}[b]
	&\left\lbrace 
	\partial_{\eta} 
	-
	\frac{D}{2}
	\nabla ^2
	+
	\frac{i}{2}
	\left[  \rcl\left(\rb,t+\frac{\eta}{2}\right) -\rcl\left(\rb,t-\frac{\eta}{2}\right) \right] 
	\right\rbrace 
	\\
	& \times
	C_{\eta,\eta'}^{t}(\rb,\rb')
	=\,
	\frac{1}{2}\delta(\eta-\eta')\delta(\rb-\rb').
\end{aligned}
\end{align}
Here $D$ is the diffusion constant. The variables $t$ and $\eta$ represent the average and relative time of the time-reserved paths, respectively.
$\rcl$ denotes the ``classical'' 
(versus quantum \cite{Keldysh})
component of the hydrodynamic density field, whose
interaction with the Cooperon results in phase relaxation.
The calculation of the WL correction requires averaging 
$C_{\eta,-\eta}^{t}(\rb,\rb)$ in Eq.~(\ref{eq:AAK_S}) 
over stochastic thermal fluctuations of $\rcl$ (denoted by the angular brackets). 

Following Refs.~\cite{AAK} and~\cite{AleinerBlanter}, the solution to Eq.~(\ref{eq:AAK_C}) 
can be represented in the form of a Feynman path integral~\cite{Feynman}. 
After averaging over the density fluctuations, we obtain
\bsub\label{eq:PathSol}
\begin{align}
\!\!	
&\begin{aligned}[b]\label{eq:Path}
	\braket{C_{\eta,-\eta}^{t}(\rb,\rb)}_{\rho}
	= \, &
	\frac{1}{2}
	\int_{\rb(-\eta)=\rb}^{\rb(\eta)=\rb} 
	{\cal D} \rb(\tau)
	\\
	& \times
	\exp
	\left\lbrace 
	-S_0[\rb(\tau)]-S_I[\rb(\tau)]
	\right\rbrace ,		 
\end{aligned}	
\\
\!\!	
&\begin{aligned}\label{eq:S0}
	S_0[\rb(\tau)]
	=\,
	\int_{-\eta}^{\eta}
	d\tau 
	\frac{1}{2D} \dot{\rb}^2(\tau)
	,
\end{aligned}	
	\\
\!\!	
&\begin{aligned}[b]\label{eq:SI}
	S_I[\rb(\tau)]
	=\,&
	\frac{1}{4} 
	\int_{-\eta}^{\eta}
	d\tau_1
	\int_{-\eta}^{\eta}
	d\tau_2
	\\
	& \times
	\!\!\!
	\left[
	\begin{aligned}
	& i \bd_{\rho}^{(K)}
	\left( \rb(\tau_1)-\rb(\tau_2),\frac{\tau_1-\tau_2}{2}\right) 
	\\
	-&i \bd_{\rho}^{(K)}
	\left( \rb(\tau_1)-\rb(-\tau_2),\frac{\tau_1-\tau_2}{2}\right) 	 
	\end{aligned}	
	\right]\!\!,\!\!		 
\end{aligned}	
\end{align}
\esub
where $i \bd_{\rho}^{(K)}$ is the Keldysh correlation function of $\rcl$. 
Notice that the dependence of the Cooperon on the average time $t$ is removed by the fluctuation average.

For long range Coulomb interactions~\cite{AAK,AleinerBlanter}, 
the noise kernel $i \bd_{\rho}^{(K)}$ in Eq.~(\ref{eq:SI}) is instantaneous in time, 
and thus of Markovian type.
By contrast, the kernel $\bd_{\rho}^{(K)}$ for short-range interactions is diffusive and non-Markovian. 
At temperature $T$, it is given by \cite{Zala-Deph,Keldysh}
\begin{align}\label{eq:Gcl}
\begin{aligned}
	i \Delta_{\rho}^{(K)} (\rb,\tau)
	\simeq\, &
	k_{\msf{B}} T  
	\frac{\gamma^2}{\kappa}
	\left( \frac{1}{4 \pi D_c \lvert \tau \rvert } \right) 
	\exp \left( {-\frac{\rb^2}{4 D_c \lvert \tau \rvert }} \right),\!\!\!\!\!\! 
\end{aligned}	
\end{align}
where $\gamma$ indicates the short-range interaction strength, 
$\kappa$ is the charge compressibility, and $D_c= D/(1-\gamma)$ is the charge diffusion constant. 

Eq.~(\ref{eq:PathSol}) can be interpreted as the path integral of a self-interacting polymer loop with boundary condition $\rb(-\eta)=\rb(\eta)=\rb$.
The Gaussian term $S_0$ in Eq.~(\ref{eq:S0}) describes the unperturbed random walk.
The interaction term $S_I$ in Eq.~(\ref{eq:SI}) consists of two parts: the repulsion between $\rb(\tau_1)$ and $\rb(\tau_2)$, 
and the attraction between $\rb(\tau_1)$ and $\rb(-\tau_2)$. 
The repulsive (attractive) interaction arises due to causal (anticausal) correlations within a path (between time-reversed paths). 
The ranges of both types of interactions are $\sqrt{D_c \lvert \tau_1 - \tau_2 \rvert}$.
Fig.~\ref{fig:F0} shows a schematic illustration of this polymer loop, where the repulsive and attractive interactions are indicated by the red dotted and blue dashed lines, respectively.
The characteristic length scale of the loop corresponds to the dephasing length $L_{\phi} = \sqrt{D\tau_{\phi}}$. 
	
Another self-interacting polymer model described by Eq.~(\ref{eq:PathSol}) is the self-avoiding random walk (SAW), 
where $S_I$ instead acquires the form
\begin{align}\label{eq:SAW}
\begin{aligned}
	&S_I[\rb(\tau)]
	=\, 
	g
	\int
	d\tau_1
	\int
	d\tau_2
	\,
	\delta \left( \rb(\tau_1)-\rb(\tau_2)\right),	 	 
\end{aligned}	
\end{align}
with $g$ the interaction constant. 
While the SAW interaction in Eq.~(\ref{eq:SAW}) is local in space (a contact interaction),
it is entirely nonlocal in ``time,'' while the interaction in Eq.~(\ref{eq:Gcl})
is nonlocal in space and time.


\textit{Replica approach}.---
It is well-known that the critical scaling of the SAW can be studied using a replica
approach \cite{CardyBook}. 
We apply a similar strategy for the dephasing problem by defining the replica field theory, 
\bsub\label{eq:FieldIntegral}	
\begin{align}
\!\!
&\begin{aligned}\label{eq:Z}	 
	Z =\,
	\int \dd \bPsi \dd \Psi \dd \rcl \,
	 e^{-S_{\Psi} [\bPsi,\Psi] - S_{\rho}[\rcl]-S_c[\bPsi,\Psi,\rcl] },
\end{aligned}	 
	\\
\!\!
&\begin{aligned}\label{eq:SPsi}
	S_{\Psi}[\bPsi,\Psi]
	=\,
	\intl{\kb,\ww}
	\bPsi^a (\kb,\ww)
	\left( 
		\xi k^2
		-i h \ww
		+r
	\right)  
	\Psi^a (\kb,\ww),
\end{aligned}
\!	 
	\\
\!\!
&\begin{aligned}\label{eq:Srho}
	S_{\rho}[\rcl]
	=\,	
	 \frac{1}{2}
	\intl{\kb,\ww}
	\rcl(\kb,\ww)
	\rcl(-\kb,-\ww)
	\,
	\frac{ k^4 + \ww^2}{2  k^2},
\end{aligned}	
	\\	 
\!\!
&\begin{aligned}[b]\label{eq:Sc}
	S_{c}[\bPsi,\Psi,\rcl]	
	=\,&
	 \frac{i}{2} \lb
	\intl{\qb, \Omega}\intl{\kb,\ww}
	\rcl(\qb, \Omega)
	\\
	& \times
	\!\!\!
	\left[ 
	\begin{aligned}
		&\bPsi^a\left(\kb+\qb,\ww+\frac{\Omega}{2}\right) 
		\\
		-&
		\bPsi^a\left(\kb+\qb,\ww-\frac{\Omega}{2}\right) 
	\end{aligned}
	\right]
	\!\!
	\Psi^a (\kb,\ww),
	\!\!
\end{aligned}
\end{align}
\esub
where $\lb$ and $\xi$ are (in $d=2$) defined by
\begin{align}\label{eq:lb}
	\lb \equiv \sqrt{k_B T \gamma^2 / \kappa D_c}, 
	\qquad
	\xi \equiv D / 2 D_c.
\end{align}
The Cooperon is encoded by the (bosonic or fermionic) field $\Psi^a$; 
the superscript $a \in \left\lbrace 1,2,...N\right\rbrace $ indexes the replica space.
The fluctuation-averaged Cooperon $\braket{C}_{\rho}$ can be obtained as the correlation function 
$	\tilde{C}(\kb,\ww) \equiv 
\braket{\Psi^a(\kb,\ww)\bPsi^a(\kb,\ww)}$ 
in the replica limit $N \rightarrow 0$,
	$\braket{C}_{\rho}(\kb,\ww)
	=
	\frac{1}{2} \tilde{C}(\sqrt{D_c}\kb,\ww). $
In Eq.~(\ref{eq:SPsi}), 
$h$ denotes the scaling factor of the frequency (which will acquire anomalous corrections), while
the ``mass term'' $r$ is not present in the bare action, but will be generated by the RG transformation
described below.

One can integrate out the density field $\rcl$ in Eq.~(\ref{eq:FieldIntegral}), and introduce a matrix field 
$\hat{Q}$ to decouple the generated quartic interaction terms. 
After integrating out $\Psi$, one arrives at an effective field theory for $\hat{Q}$, 
whose saddle point $\QSP$ gives the SCBA dephasing rate (in $d = 2$)~\cite{Zala-Deph,SuppMaterial}:
\begin{align}\label{eq:SCBA}
\begin{aligned}[b]
	\tau^{-1}_{\msf{SCBA}}
	=\,&
	2
	\intl{\kb,\ww}	
	\left[ 
		D k^2 
		-
		i \ww
		+
		\tau^{-1}_{\msf{SCBA}}
	\right]^{-1}
	\left[ i \bd_\rho^{(K)}(\kb,\ww)  \right]
	\\
	=\,&
	\frac{1}{2 \pi D_c \kappa}
	\frac{\gamma^2}{\left(2-\gamma\right)}
	\,
	k_{\msf{B}} T 
	\ln \left( \frac{k_{\msf{B}} T}{\tau_{\msf{SCBA}}^{-1}}\right).			
\end{aligned}
\end{align}
This result is identical to that obtained from the lowest order cumulant expansion of Eq.~(\ref{eq:Path}), 
when the infrared limit of the integral is cut off at $\tau^{-1}_{\msf{SCBA}}$ ``by hand''~\cite{ChakravartySchmid}; 
the result also obtains via self-consistent diagrammatic perturbation theory~\cite{Zala-Deph, AAG, Keldysh}. 
The SCBA is exact for long-range screened Coulomb interactions~\cite{AAK}, but there can be
corrections for a correlated (non-Markovian) bath. 

We return to the replicated field integral in Eq.~(\ref{eq:FieldIntegral}).
In $d$ spatial dimensions, the engineering dimensions of the fields and coupling constants are
	$\left[ \Psi (\kb,\ww)\right]  
	=\,
	\left[ \rcl (\kb,\ww) \right]
	=\,
	 - 2 - d/2$, 
	$\left[ \lb \right]  
	=\,
	(4 - d)/2$, 
	$\left[ r \right] 
	= \,
	2$,
and
	$\left[ \xi \right] 
	=
	\left[ h \right]  
	= 0$.
Here we have adopted the convention that momentum carries dimension 1, i.e.\ $[\kb]=1$,
and frequency carries the engineering dimension $z = 2$, $[\omega] = 2$.  
The upper critical dimension at which coupling constant $\lb$ becomes dimensionless is $d_c=4$. 
Below (above) $d_c$, the interaction term $S_c$ in Eq.~(\ref{eq:Sc}) is relevant (irrelevant)
in a renormalization group (RG) sense. 
This suggests that, below $d_c$ spatial dimensions, there might exist a nontrivial fixed point 
that is perturbatively accessible in a controlled $\epsilon$-expansion from $d_c$.

Besides its dephasing effect, the density fluctuation $\rcl$ also contributes Altshuler-Aronov corrections 
to the conductivity \cite{AA,Keldysh}; these are ignored here.
Moreover, since the weak localization correction in Eq.~(\ref{eq:AAK_S}) is itself the virtual shift of the
diffusion constant $D$, we assume that $D$ does not change under the RG flow
\emph{within the dephasing problem}.  
In particular, we employ the renormalization scheme where, contrary to frequency scaling factor $h$ 
in Eq.~(\ref{eq:SPsi})
that flows under the
RG transformation, $\xi$ defined in Eq.~(\ref{eq:lb}) is fixed by a wave function renormalization.
A more sophisticated approach might impose a scale-dependent
self-consistent condition on $D$, as it enters both the Cooperon and bath correlation functions
that together determine the weak localization correction up to that scale.

\begin{figure}[b!]
	\subfigure{\includegraphics[angle=0,width=.333\textwidth]{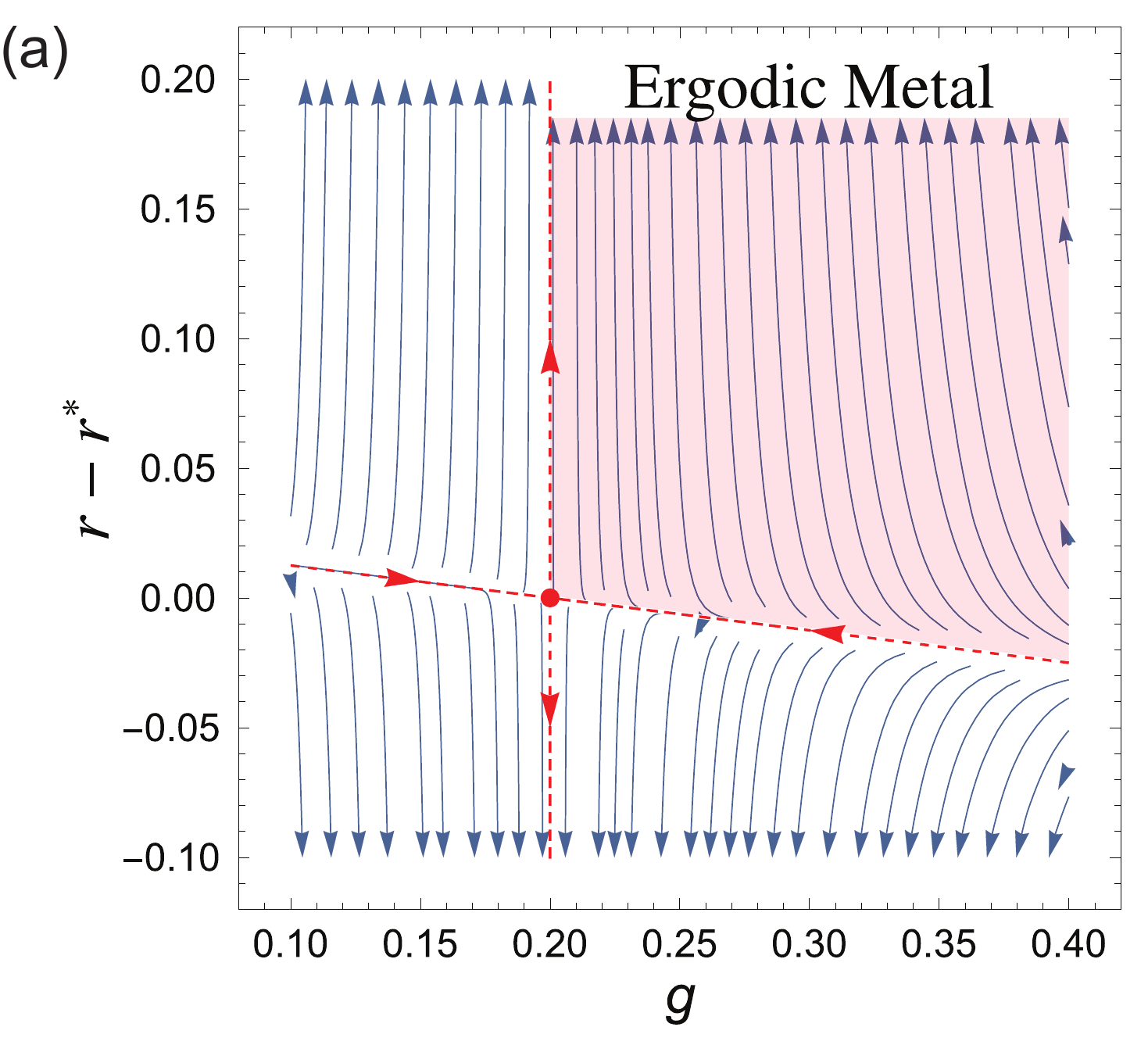}}
	\\
	\vspace{-12pt}
	\subfigure{\includegraphics[angle=0,width=.333\textwidth]{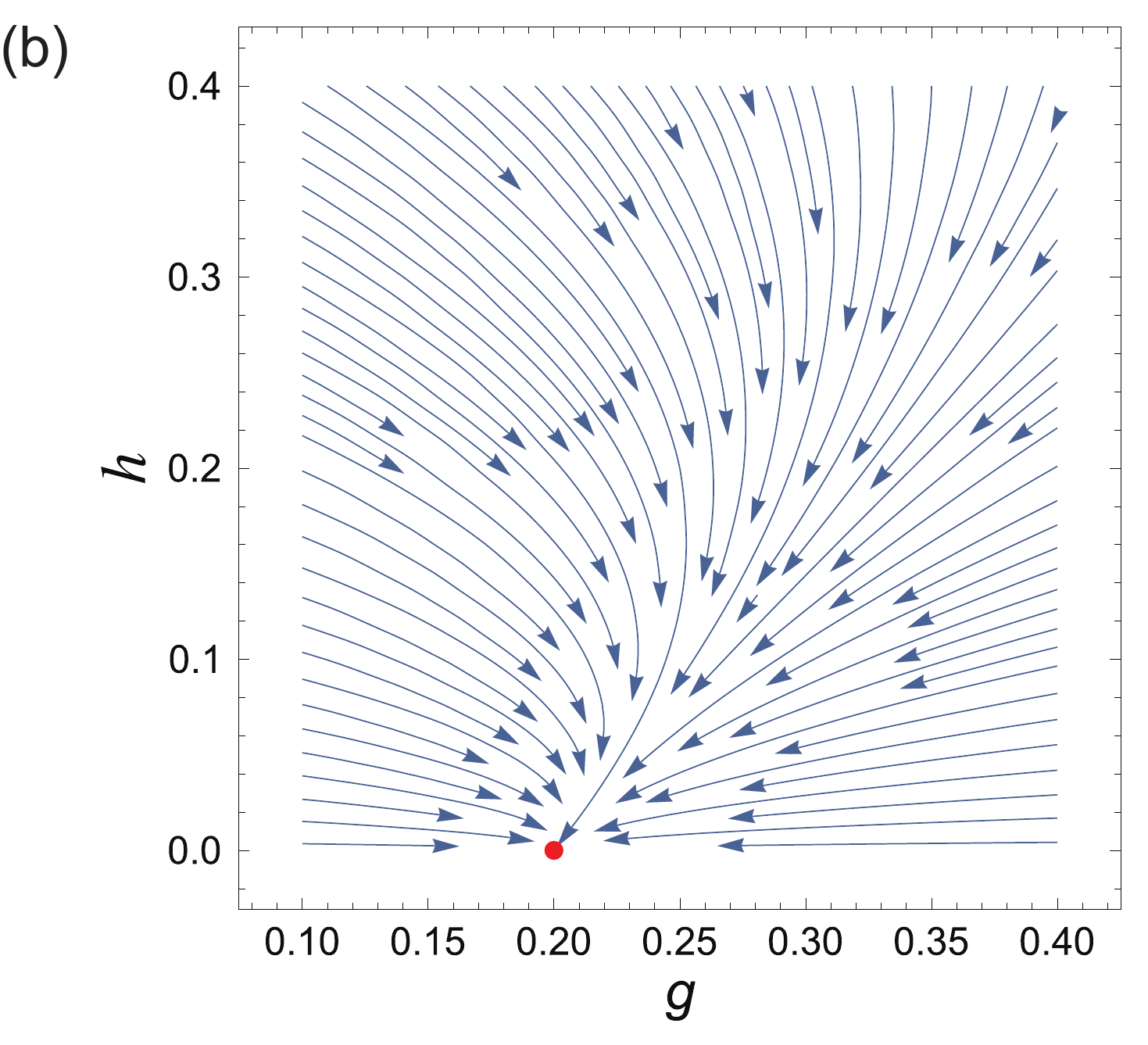}}
	\vspace{-12pt}
	\caption{
	One-loop RG flow in spatial dimensions $d=4-\epsilon$.
	Panel (a) depicts the RG flow in the $g-r$ plane with $h=0$, $\xi=1$ and $\epsilon=0.1$. 
	Here the coupling constant $g$ is proportional to $k_B T/D$ (in $d=2$). 
	The nontrivial fixed point denoted by the red dot has two irrelevant and one relevant directions. 
	(The relevant direction and one irrelevant direction lie in the $g-r$ plane and are indicated by the red dashed lines.)
	The dephasing rate $\tau_{\phi}^{-1}=2(r-r^*)$ vanishes at the critical fixed point corresponding to temperature $T^*>0$.
	Above $T^*$, the system flows towards the ergodic metal phase where the dephasing rate $\tau_{\phi}^{-1}$ is nonzero.
	The RG analysis suggests that the system undergoes a temperature-tuned transition that can be considered 
	as a possible instability of the ergodic phase (in $d = 2$). 
	Panel (b) shows the RG flow in the $g-h$ plane.
	}
	\label{fig:F1,2}
\end{figure}

Carrying out a Wilsonian RG analysis in $d=4-\epsilon$ dimensions at one-loop level, 
we obtain the $\beta$-functions to leading order in $\epsilon$
\cite{SuppMaterial}:
\begin{align}\label{eq:beta}
\begin{aligned}
		&\frac{d g}{dl} 
		=\,
		\epsilon g
		-g^2\frac{4\xi}{(h+2\xi)^3}, 
		\\
		&\frac{d r}{dl} 
		=\,
		2 r 
		-g r \frac{2\xi}{(h+2\xi)^3}
		+\frac{g}{2}\frac{1}{(h+2\xi)},	
		\\
		&\frac{d h}{d l} 
		=\,
		-g \frac{2\xi}{(h+2\xi)^3}h,		
\end{aligned}
\end{align}
where $l$ is the logarithm of the length scale, and $g$ is defined by 
$g \equiv \lb^2/(4\pi^2)$.
These equations possess a nontrivial fixed point at 
    $g^* = \, 2 \xi^2 \epsilon$,
	$r^* =\,  -\epsilon \xi/4$,
	$h^* = \, 0. $	    

In Figs.~\ref{fig:F1,2}(a) and (b), we show the RG flow described by Eq.~(\ref{eq:beta}) in the $g-r$ ($h=0$) and $g-h$ planes, respectively. 
We have set $\xi=1$ and $\epsilon=0.1$ in both plots.
The red dashed line slightly tilted from the horizontal $g$-axis in Fig.~\ref{fig:F1,2}(a) represents the critical surface in the $h=0$ plane.
Under the RG transformation, any point on the critical surface flows towards the nontrivial fixed point (denoted by the red dot) corresponding to $T^*>0$
(since $g \propto T$). 
The dephasing rate $\tau_{\phi}^{-1}=2(r-r^*)$ vanishes at this critical point. 
By contrast, for those points off of the critical surface and associated with $T>T^*$ [shaded region in Fig.~\ref{fig:F1,2}(a)], 
the RG flow is directed away from the critical fixed point and towards the ergodic phase where a nonzero dephasing rate $\tau_{\phi}^{-1}$ 
is generated, as in the SCBA [Eq.~(\ref{eq:SCBA})].
We only consider the critical point as approached from the ergodic phase,
so we do not consider parts of the flow outside
the critical surface or shaded region. 
Fig.~\ref{fig:F1,2}(b) shows that the frequency scale $h$ flows to zero at the fixed point. 

At the critical fixed point, the Cooperon has the scaling behavior  
$\braket{C}_{\rho}\left( \kb, \ww \rightarrow 0 \right)
	=\,
	1/D k^2,$
which is the bare (undephased) result \cite{SuppMaterial}. 
Inserting this into Eq.~(\ref{eq:AAK_S}) gives a logarithmic divergence of the weak localization correction in the infrared in 2D. 
By linearizing the RG flow near the nontrivial fixed point \cite{SuppMaterial}, we find that in spatial dimensions $d=4-\epsilon$, 
the coupling constant $(r - r^*)$ has the scaling dimension $y_r= 2-\epsilon/2+ O(\epsilon^2)$,
which leads to a correlation length exponent $\nu=1/2+\epsilon/8 + O(\epsilon^2)$. 
We note that the critical exponent $\nu=2/d$, which saturates the Harris-Chayes bound ($\nu \geq 2/d$)~\cite{Harris, Chayes}.

As emphasized above, our result gives only a hint of a possible 
instability of the ergodic phase in
$d = 2$ dimensions. 
A challenging but worthwhile direction for future work is to impose a scale-dependent self-consistency condition
between the RG controlling the dephasing due to real processes, and the running renormalization of the 
diffusion constant due to the virtual ones. This self-consistent \cite{BAA,SG-2} (``RG-improved perturbation theory'') 
approach would account for the fact that the diffusion constant itself changes with scale. 
At the transition in 2D, it may reach a critical (universal) average value or vanish altogether \cite{BAA,Gornyi05,MBL1DT1,MBL1DT2,MBL1DT3}. 
A further refinement would incorporate higher order quantum interference conductance and Altshuler-Aronov corrections. 

We thank 
Bitan Roy,
Arijeet Pal,
Sarang Gopalakrishnan,
and David Huse for helpful discussions. 
This work was supported by the Welch Foundation Grant No.~C-1809 and by NSF CAREER Grant No.~DMR-1552327.
MSF thanks the Aspen Center for Physics, which is supported
by the NSF Grant No.~PHY-1066293, for its hospitality while part of this work was performed.

\newpage \clearpage

\onecolumngrid

\begin{center}
	{\Large
	Dephasing catastrophe in $4 - \epsilon$ dimensions:\\
	A possible instability of the ergodic (many-body-delocalized) phase
	\vspace{4pt}
	\\
	SUPPLEMENTAL MATERIAL
	}
\end{center}
\makeatletter
\setcounter{equation}{0}
\setcounter{figure}{0}
\setcounter{table}{0}
\setcounter{page}{1}
\renewcommand{\theequation}{S\arabic{equation}}
\renewcommand{\thefigure}{S\arabic{figure}}
\renewcommand{\bibnumfmt}[1]{[S#1]}
\renewcommand{\citenumfont}[1]{S#1}


\section{I.\ Anderson delocalization transitions in 1D versus higher dimensions $d \geq 2$}

We emphasize that most work performed on interacting, disordered quantum many-body systems
since \cite{S-BAA} has focused on one-dimensional spin systems, see e.g.\ 
\cite{S-MBL-Rev,S-Imbrie2016,S-MBL1DT1,S-MBL1DT2,S-MBL1DT3,S-RH2017}.
This paper concerns a system of fermions in $d > 1$. 
If many-body mobility edges can exist in such a system (as claimed by \cite{S-BAA}), 
then one can imagine deforming a finite-temperature ergodic-to-MBL transition to 
a \emph{zero-temperature} metal-insulator transition (MIT). For a particular scenario, see \cite{S-Keldysh}. 
It is important to emphasize that there exists a well-established phenomenology of \emph{zero-temperature} MITs
and/or critical delocalization in $d = 2,3$ that 
is {\bf completely different from 1D}, where delocalization only occurs at zero temperature via infinite randomness fixed 
points \cite{S-Fisher94,S-BalentsFisher97,S-Mathur97,S-Vasseur16}.
In particular, these traits include 
\begin{enumerate}
	\item[(a)]{a finite dynamic critical exponent,} 
	\item[(b)]{weak (as opposed to ``frozen'' \cite{S-Foster2009,S-BalentsFisher97}) multifractality, and} 
	\item[(c)]{log-normal (not Pareto) tails for distribution functions of the local density of states or conductance \cite{S-AKL1991,S-Mathur97}.} 
\end{enumerate}
These traits appear e.g.\ in exactly solved 2D conformal field theories 
\cite{S-Mudry1996,S-Caux1996},
quantum Hall plateau transitions 
\cite{S-Huckestein1995,S-Evers2003,S-Mirlin2003},
the MIT in the spin-orbit class \cite{S-AndRev}, 
and
noninteracting Anderson MITs in $2 + \epsilon$ \cite{S-AndRev} and three dimensions \cite{S-Rodriguez2009}.
In the orthogonal class, the features (a)--(c) already appear at lowest order in perturbation theory
\cite{S-AKL1991}.
While only a few of these systems are well-understood when interactions are also included
\cite{S-BK94,S-FinkelRev,S-QHInt,S-Foster2014}, 
the phenomenology does not change \cite{S-BK94,S-MFCCoul,S-MFCCoul2,S-Foster2014}
unless the interactions completely destroy the transition and/or metallic phase \cite{S-BK94}.


\section{II.\ Saddle point and SCBA}

Here we show the derivation of the SCBA dephasing rate from the saddle point of an effective matrix field theory.
The starting point is the functional integral in Eq.~(6). After rescaling
$ 
\Psi(\kb,\ww) \rightarrow  \Psi(\kb/\sqrt{D_c},\ww)/\sqrt{D_c}, 	
\rcl(\kb,\ww) \rightarrow  \rcl(\kb/\sqrt{D_c},\ww)/\sqrt{D_c\alpha},
$
the bare action (with $r=0,h=1$) acquires the form
\begin{align}
\begin{aligned}
	S [\bPsi,\Psi,\rcl]
	=\,&
	\intl{\eta,\rb}
	\bPsi^a (\rb,\eta)
	\left(  
	\partial_{\eta} 
	-
	\frac{D}{2}
	\nabla ^2
	\right) 
	\Psi^a (\rb,\eta)	 
	+ \frac{1}{2}
	\intl{\kb,\ww}
	\rcl(\kb,\ww)
	\rcl(-\kb,-\ww)
	\left[ i \bd_{\rho}^{(K)}(\kb,\ww) \right]^{-1} 	
	\\	 
	& + i
	\intl{\eta,\rb}
	\rcl(\rb, \eta)
	\left[ 
	\bPsi^a (\rb,2\eta) \Psi^a (\rb,2\eta)
	-
	\bPsi^a (\rb,-2\eta) \Psi^a (\rb,-2\eta)
	\right] .
\end{aligned}
\end{align}
Note that now we have $\braket{C}_\rho(\kb,\ww)= \frac{1}{2} \tilde{C}(\kb,\ww)$.

By performing the average over the density fluctuation $\rcl$, 
one arrives at two effective quartic interaction terms:
\begin{align}\label{Seq:I}
\begin{aligned}
	S_I[\bPsi,\Psi]
	=\,
	& 
	\frac{1}{4}	
	\intl{\eta,\rb,\eta',\rb'}
	\left[ 
		i\bd_{\rho}^{(K)}\left(\rb-\rb',\frac{\eta-\eta'}{2}\right)  
		-
		i\bd_{\rho}^{(K)}\left(\rb-\rb',\frac{\eta+\eta'}{2}\right)
	\right]
	\bPsi^a (\rb,\eta) \Psi^a (\rb,\eta)
	\bPsi^b (\rb',\eta') \Psi^b (\rb',\eta').	
\end{aligned}
\end{align}
In the following calculation, we drop the second term in Eq.~(\ref{Seq:I}), since
it is expected to give a less singular contribution \cite{S-Keldysh}. 
We also take $\Psi$ to be a fermionic field. 
The interaction term is then decoupled by a matrix field $\hat{Q}^{ab}(\rb,\rb';\eta,\eta')$ 
that shares the same structure with
$\Psi^a (\rb,\eta) \bPsi^b (\rb',\eta')$,	
\begin{align}
\begin{aligned}
	Z
	=\, &
	\int \dd \hat{Q} 
	\exp 
	\left\lbrace 
	-\frac{1}{2}	
	\intl{\eta,\rb,\eta',\rb'}
	\hat{Q}^{ab}(\rb,\rb';\eta,\eta')
	\left[ i\bd_{\rho}^{(K)}\left(\rb-\rb',\frac{\eta-\eta'}{2}\right)  \right]^{-1}
	\hat{Q}^{ba}(\rb',\rb;\eta',\eta)
	\right\rbrace 
	\\
	& \times
	\int \dd \bPsi \dd \Psi
	\exp 
	\left\lbrace 
	-	
	\intl{\eta,\rb,\eta',\rb'}
	\bPsi^a (\rb,\eta)
	\left[ 
	\left( 
	\partial_{\eta} 
	-
	\frac{D}{2}
	\nabla^2
	\right)  
	\delta^{ab}
	\delta (\rb-\rb')
	\delta (\eta-\eta')
	+
	\frac{1}{\sqrt{2}}	
	\hat{Q}^{ab}(\rb,\rb';\eta,\eta')
	\right] 
	\Psi^b (\rb',\eta')
	\right\rbrace 	.
\end{aligned}
\end{align}
Integrating out the fermionic field $\Psi$ leads to an effective field theory for the matrix $\hat{Q}$:
\begin{align}
\begin{aligned}
	Z =\, &
	\int \dd \hat{Q}
	\exp
	\left\lbrace 
		-\frac{1}{2}	
		\intl{\eta,\rb,\eta',\rb'}
		\hat{Q}^{ab}(\rb,\rb';\eta,\eta')
		\left[ i\bd_{\rho}^{(K)}\left(\rb-\rb',\frac{\eta-\eta'}{2}\right)  \right]^{-1}
		\hat{Q}^{ba}(\rb',\rb;\eta',\eta)
		 +
		\Tr \ln	
		\left[ 
			\partial_{\eta} 
			-
			\frac{D}{2}
			\nabla^2
			+
			\frac{1}{\sqrt{2}}	
			\hat{Q}
		\right] 
	\right\rbrace 	.
\end{aligned}
\end{align} 
Taking variation of the action with respect to $\hat{Q}$, one obtains the saddle point equation:
\begin{align}\label{eq:SPEQc}
\begin{aligned}
	& 
	\hat{Q}^{ba}(\kb',\kb;\ww',\ww)
	=\,
	\intl{\bl,\nu}
	\frac{1}{\sqrt{2}}	
	\bra{a,\kb+\bl,\ww+\frac{\nu}{2}}
	\left\{ 
	\left[
	-i \left(\ww+\frac{\nu}{2}\right) 
	+
	\frac{D}{2}
	(\kb+\bl)^2
	\right]  
	+
	\frac{1}{\sqrt{2}}	
	\hat{Q}
	\right\}^{-1}	
	\ket{b,\kb'+\bl,\ww'+\frac{\nu}{2}}	
	\left[ i\bd_{\rho}^{(K)}(\bl,\nu) \right] .
\end{aligned}
\end{align}
We then look for a static, replica-symmetric, translationally invariant solution of the form 
$
\hat{Q}^{ab}(\kb,\kb';\ww,\ww') 
= 
(2\pi)^d\delta (\kb-\kb') 2\pi \delta (\ww-\ww')
\hat{I}^{ab}
\QSP
$
and find
\begin{align}
\begin{aligned}
	\QSP
	=\,
	\frac{1}{\sqrt{2}}
	\intl{\vex{l},\nu}	
	\left[ 
	\left( 
	-i \frac{\nu}{2}
	+
	\frac{D}{2}
	l^2
	\right)  
	+
	\frac{1}{\sqrt{2}}	
	\QSP
	\right]^{-1}
	\left[ i \bd_{\rho}^{(K)}(\vex{l},\nu)  \right].		
\end{aligned}
\end{align}
The averaged Cooperon $\braket{C}_{\rho}(\kb,\ww)$ now acquires a SCBA dephasing rate given by the saddle point $\tau^{-1}_{\msf{SCBA}}=\sqrt{2}\QSP$ 
[see Eq.~(8)]:
\begin{align}
\begin{aligned}
	\braket{C}_{\rho}(\kb,\ww)
	=\,
	\frac{1}
	{
		D k^2
		-2 i \ww 
		+\tau^{-1}_{\msf{SCBA}}	
	}.
\end{aligned}	 
\end{align}


\section{III.\ 1-loop field theory renormalization group}

\begin{figure}[t!]
	\includegraphics[angle=0,width=.6\textwidth]{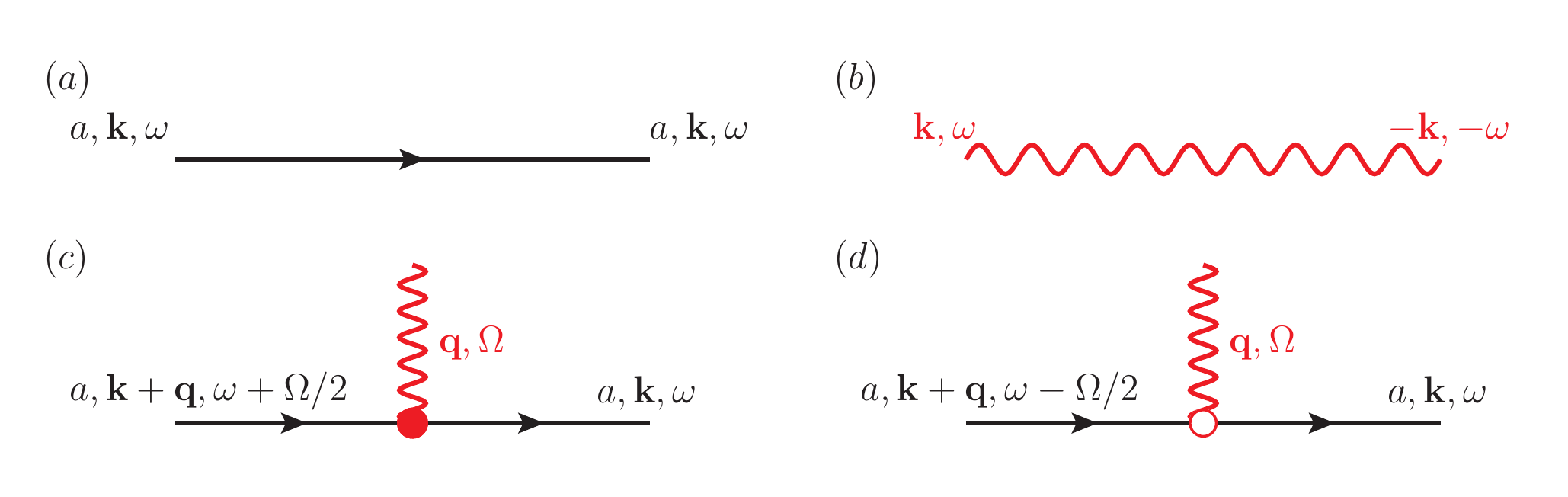}
	\caption{Feynman rules: Diagrams in (a) and (b) represent the bare propagators for fields $\Psi$ and $\rho$, respectively. 
	(c) and (d) show the interaction vertices coupling together these two fields. 
	}
	\label{Sfig:SF1}
\end{figure}

In Fig.~\ref{Sfig:SF1}, we show the Feynman rules for the replica field theory in Eq.~(6). 
Here $\Psi$ and $\rcl$ are represented diagrammatically by (black) solid and (red) curvy lines, respectively.
Their bare propagators are given by  
\bsub\label{eq:Propagator}
\begin{align}
&\begin{aligned}
	\braket{\Psi^a(\kb,\ww)\bPsi^b(\kb,\ww)}_0
	=\,
	\delta^{a,b} \bd(k,\ww)
	=\,
	\frac{\delta^{a,b}}{\xi k^2 - i h \ww + r},
\end{aligned}
\\
&\begin{aligned}
	\braket{\rcl(\kb,\ww)\rcl(-\kb,-\ww)}
	=\,
	\Gcl(k,\ww)
	=\,
	\frac{2 k^2}{ k^4 + \ww^2 },
\end{aligned}
\end{align}
\esub
and are depicted in Figs.~\ref{Sfig:SF1}(a) and~\ref{Sfig:SF1}(b).
Figs.~\ref{Sfig:SF1}(c) and~\ref{Sfig:SF1}(d) illustrate the interaction vertices coupling between $\Psi$ and $\rcl$ fields. 
They arise from the first [diagram~(c)] and second term [diagram~(d)] in the interaction action $S_c$
[Eq.~(6d)] and are named, respectively, causal and anticausal interaction vertices in the following,
with amplitudes $\mp i \lambda/2$.  
We use a solid (open) dot to represent the causal (anticausal) vertex.

The field theory contains coupling constants: $\lb$, $r$, $h$ and $\xi$. We employ the RG scheme where $\lb$, $r$ and $h$ flow under the RG transformation, while $\xi$ is fixed by wave function renormalization.

\begin{figure}[t!]
	\includegraphics[angle=0,width=.8\textwidth]{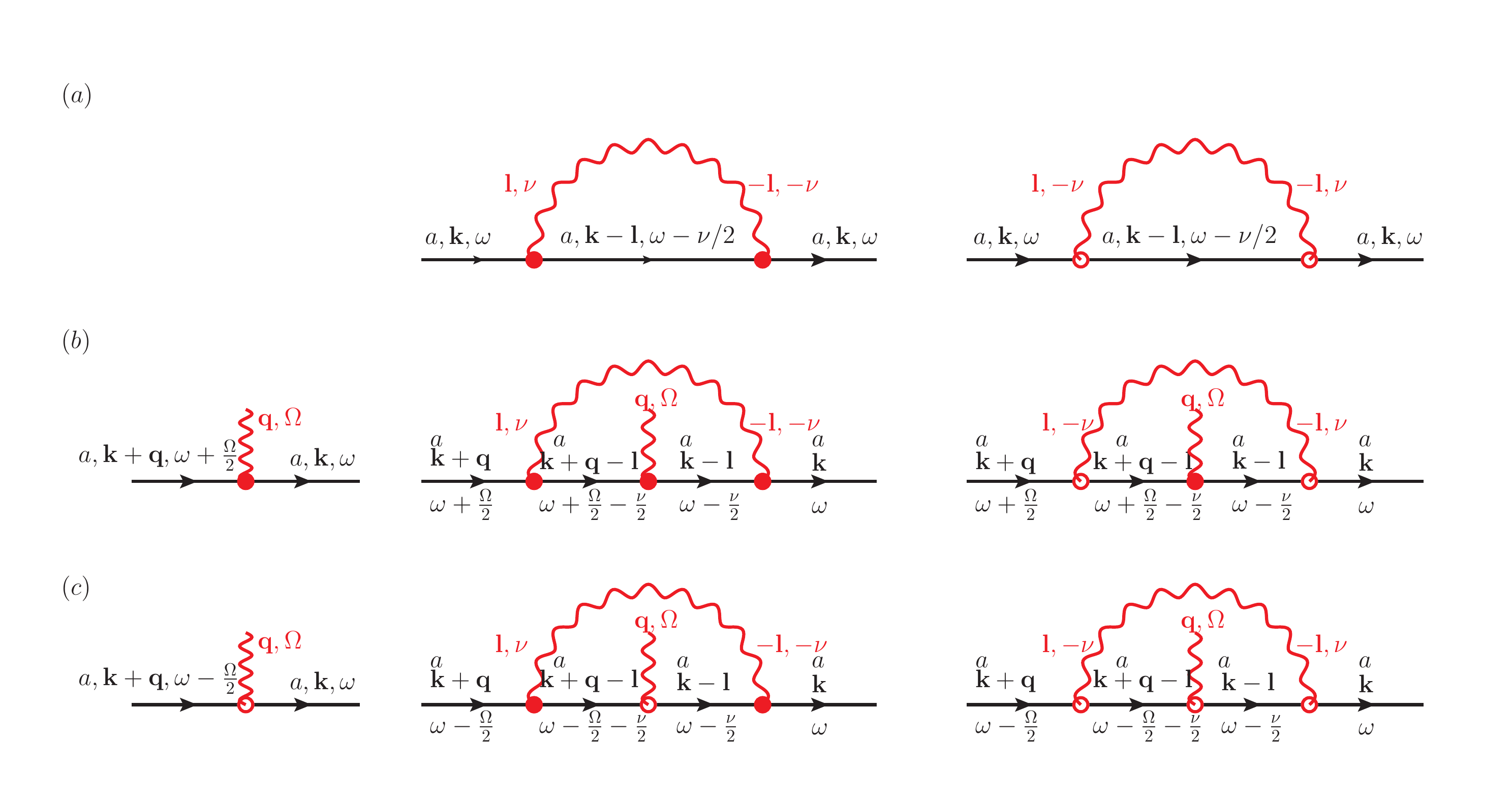}
	\caption{One-loop diagrams that contribute to the renormalization of the coupling constants: 
	(a) depicts diagrams of self-energies diagonal in frequency space. 
	(b) and (c) represent the three-point irreducible vertex corrections to the causal and anti-causal vertices.
	Vertex corrections are neglected in the SCBA calculation, presented in the previous section and yielding Eq.~(8). 
	}
	\label{Sfig:SF2}
\end{figure}

Diagrams in Fig.~\ref{Sfig:SF2}(a) represent the (ultraviolet divergent) 
$\Psi$-field self-energy contributions diagonal in frequency space, and contribute to the renormalization of $h$ and $r$ 
to one-loop order.
The renormalization of $\lb$ is determined by diagrams in Figs.~\ref{Sfig:SF2}(b) and~\ref{Sfig:SF2}(c), 
which show the three-point irreducible vertex functions corresponding to causal (C) and anticausal (AC) vertices, respectively.
Note that diagrams with closed $\Psi$ loops do not contribute in the replica limit $N \rightarrow 0$, 
and are not included here.

The corresponding two-point and three-point bare irreducible vertex functions 
respectively 
assume the forms
\bsub
\begin{align}
&\begin{aligned}\label{Seq:Gamma2}
	\Gamma^{(0)}_{2 \Psi,0\rcl}(\kb,\ww)
	=\,
	-\bd^{-1}(\kb,\ww)
	+	
	2\left( \frac{i\lb}{2}\right) ^2	 
	\intl{\bl,\nu}
	\bd\left( \Avert{\kb-\bl},\ww-\frac{\nu}{2}\right)  
	\Gcl(l,\nu),
\end{aligned}	
	\\
&\begin{aligned}\label{Seq:Gamma3}
	\Gamma^{(0)}_{2 \Psi, 1\rcl(C/AC)}(\kb,\ww;\qb,\Omega)
	=\,
	\mp \frac{ i\lb}{2}
	\left[ 
	1
	+	
	2\left( \frac{i\lb}{2}\right) ^2	 
	\intl{\bl,\nu}
	\bd \left( \Avert{\kb+\qb -\bl},\ww \pm \frac{\Omega}{2}-\frac{\nu}{2}\right)  
	\bd \left( \Avert{\kb-\bl},\ww-\frac{\nu}{2}\right)  
	\Gcl(l,\nu)
	\right] .
\end{aligned}	 
\end{align}
\esub
Upon evaluating these integrals, and keeping only the relevant terms, we find that to the lowest order in $\epsilon$
\begin{align}
	\begin{aligned}[b]
	\Gamma^{(0)}_{2 \Psi,0\rcl}(\kb,\ww)
	=&\,
	-
	\left[ \xi k^2 - i h \ww + r \right] 
	-
	\frac{\lb^2}{16\pi^2}\frac{1}{h+2\xi}\Lambda^2
	-
	\left[ \xi k^2 - i h \ww + r \right] 
	\left[ -\frac{\lb^2}{4\pi^2}\frac{1}{(h+2\xi)^2} \ln \Lambda\right] 
\\&\,
	- 
	\xi k^2
	\left[ 	   \frac{\lb^2}{4\pi^2}\frac{2\xi}{(h+2\xi)^3} \ln \Lambda \right],
\\
	\Gamma^{(0)}_{2 \Psi, 1\rcl(C/AC)}(\kb,\ww;\qb,\Omega)
	=&\,
	\mp \frac{ i\lb}{2}
	\left[ 
	1
	-
	\frac{\lb^2}{4\pi^2}\frac{1}{(h+2\xi)^2} \ln \Lambda
	\right]. 	  		  
\end{aligned}
\end{align}   
Here $\Lambda$ denotes the ultraviolet momentum cutoff, and frequency integrations
are performed over the entire real line.

Using the renormalization conditions \cite{S-Justin}
\begin{align}
\begin{aligned}
	\frac{d}{d\ln \Lambda} \left( Z_{\Psi}\Gamma^{(0)}_{2\Psi,0\rcl}\right)  =\,0 ,
	\qquad
	\frac{d}{d\ln \Lambda} \left( Z_{\Psi}\Gamma^{(0)}_{2\Psi,1\rcl} \right)  =\,0 ,
\end{aligned}	
\end{align} 
where $Z_{\Psi}$ denotes the wave function renormalization of $\Psi$, 
and rescaling $r \rightarrow r \Lambda^2$,
one obtains 
\begin{align}\label{Seq:RG1}
\begin{aligned}
	&
	\frac{d \ln Z_{\Psi}}{d\ln \Lambda}
	=\,
	\frac{\lb^2}{4\pi^2}\frac{h}{(h+2\xi)^3} ,
	\\
	&\frac{d \ln h}{d\ln \Lambda} 
	=\,
	\frac{d \ln \lb}{d\ln \Lambda} 
	=\,
	\frac{\lb^2}{4\pi^2}\frac{2\xi}{(h+2\xi)^3},
	\\
	&\frac{d r}{d\ln \Lambda} 
	=\,
	-2 r
	+\frac{\lb^2}{4\pi^2}\frac{2\xi}{(h+2\xi)^3} r
	-\frac{\lb^2}{8\pi^2}\frac{1}{(h+2\xi)}.
\end{aligned}
\end{align}
After a change of variable [$g\equiv \lb^2/(4\pi^2)$] and dimensional analysis
(with $l \equiv -\ln\Lambda$), we obtain 
the RG equations given in Eq.~(9). 
These equations possess a nontrivial fixed point at 
    $g^* = \, 2 \xi^2 \epsilon$,
	$r^* =\,  -\epsilon \xi/4$,
	$h^* = \, 0. $	
Linearizing the $\beta$-functions around this non-trivial fixed point, we find
\begin{align}
	g(l)-g^*
	=\,&
	-8\xi z_2(0) \, e^{-\epsilon l}, 
\nonumber\\
	r(l)-r^*
	=\,&
	z_1(0) \, e^{(2-\epsilon/2) l} 
	+
	z_2(0) \, e^{-\epsilon l},
\nonumber\\
	h(l)-h^*
	=\,&
	z_3(0) \, e^{-\epsilon l /2},	
\end{align}
where the initial conditions $z_i(0), i\in\left\lbrace 1,2,3\right\rbrace $ are determined by the bare coupling constants.
Using this result, we obtain the scaling dimension of the Cooperon mass term (dephasing rate $2r -2r^*$),
$y_r = 2-\epsilon/2+O(\epsilon^2) $, 
which yields the correlation length exponent 
$\nu=1/y_r=1/2+\epsilon/8 + O(\epsilon^2)$.

We find that the Cooperon correlator 
$\tilde{C}(\kb,\ww)$ defined below Eq.~(7) in the main text
exhibits a scaling behavior of the form 
\begin{eqnarray}\label{eq:CScaling}
	 \tilde{C}\left( \kb, \ww; g(0), h(0), r(0) \right)
	 &=&
	 \exp \left[ -\int_0^l dl' \Delta_{\msf{c}}(l')\right] 	 	 
	 \tilde{C}\left[e^{l}\kb, e^{2l}\ww; g(l), h(l), r(l) \right]\!, 
	 \nonumber\\
	 \Delta_{\msf{c}}(l)
	 &=&
	 -2-  g(l) \frac{h(l)}{[h(l)+2\xi]^3},  
\end{eqnarray} 
where the first and second terms in $\Delta_{\msf{c}}$ represent the engineering and anomalous dimensions, respectively.
At the critical fixed point, the Cooperon has a scaling dimension $\Delta_{\msf{c}} \rightarrow -2$, so that 
$\braket{C}_{\rho}\left( \kb, \ww \rightarrow 0 \right)
	=\,
	1/D k^2,$
which is the bare (undephased) result.

\end{document}